\documentclass[preprint,prc,aps,showpacs,amsmath]{revtex4}

\begin{document}
\hspace{9.8 cm}FZJ-IKP-TH-2003-4
\title{An analysis of the reaction $pp\rightarrow pp\eta$ near threshold}
\author{K. Nakayama$^{a,b}$, J. Haidenbauer$^b$, C. Hanhart$^b$, and J. Speth$^b$}
\affiliation{$^a$Department of Physics and Astronomy, University of Georgia, 
Athens, GA 30602, USA \\
$^b$Institut f\"ur Kernphysik, Forschungszentrum J\"ulich,
D-52425, J\"ulich, Germany }

\begin{abstract}
It is shown that most of the available data on the $pp\rightarrow pp\eta$ reaction, 
including the invariant mass distributions in the $pp\rightarrow pp\eta$ reaction 
recently measured at COSY, can be understood in terms of the partial-wave amplitudes 
involving final $pp$ $S$ and $P$ states and the $\eta$ meson $s$-wave. This finding, 
together with the fact that results within a meson--exchange model are especially 
sensitive to the details of the excitation mechanism of the $S_{11}(1535)$ resonance, 
demonstrates the possibility of investigating the properties of this resonance in $NN$ 
collisions. The spin correlation function $C_{xx}$ is shown to disentangle the $S$- and 
$P$-wave contributions. It is also argued that spin correlations may be used to help 
constrain the contributions of the amplitudes corresponding to the final $pp$ $^3P_0$ 
and $^3P_2$ states. 
\end{abstract}

\vskip 0.5cm

\maketitle

\newpage

The primary motivations for studying the production of mesons off nucleons and nuclei are
to investigate the structure and properties of the nucleon resonances and to learn about 
hadron dynamics at short range. As far as hadron-induced reactions are concerned, and 
specifically nucleon-nucleon ($NN$) collisions, there is already a wealth of information 
on the production of the lightest meson, the pion. In particular, there now exists a 
fairly accurate and complete set of data, especially for $\pi^o$ production in the 
near-threshold energy region \cite{Meyer}, which should allow for a partial wave analysis.
The $\eta$ meson, which is the next lightest non-strange meson in the meson mass 
spectrum, has also been the subject of considerable interest. A peculiar feature of this 
meson is that it couples strongly to the $S_{11}(1535)$ nucleon resonance, which offers a
unique opportunity for investigating the properties of this resonance. Unfortunately, the
experimental information on $\eta$ production in $NN$ collisions 
\cite{EXP,Calena,Calend,Winter,Roderburg,Moskal} is much less complete than for pion 
production and is not yet sufficient for a model-independent partial-wave analysis. However, 
the available data base has greatly expanded recently thanks to measurements by the TOF and 
COSY-11 collaborations at COSY \cite{Roderburg,Moskal} that provided, not only $\eta$ and 
proton angular distributions, but also $pp$ and $p\eta$ invariant mass distributions for 
the reaction $pp\rightarrow pp\eta$. These new data, together with the earlier measurements 
\cite{EXP,Calena,Calend}, open the possibility for investigating this reaction in much more 
detail than could be done previously.

A general feature of meson production in $NN$ collisions is that the energy dependence 
of the total cross section in the near-threshold region is basically dictated by the 
available phase space plus the $NN$ final state interaction (FSI) in $S$ states. The 
effect of the strong $NN$ FSI also shows up in the corresponding $NN$ invariant mass 
spectrum as a peak close to the threshold value of the invariant mass, $m_{NN}=2m_N$, 
where $m_N$ denotes the nucleon mass. Surprisingly, the recently measured $pp$ invariant 
mass distribution in the reaction $pp\rightarrow pp\eta$ \cite{Roderburg,Moskal} 
at excess energies of $Q=15$ and $41$ MeV shows, in addition to a peak very close to the 
threshold, a broad bump at higher values of $m_{pp}$ (see Fig.~\ref{fig:mpp}). While the 
peak can easily be understood as arising from the strong $pp$ FSI in the $^1S_0$ state as 
mentioned above, it is not trivial to explain the origin of the bump at higher $m_{pp}$. 

The purpose of the present work is to analyze this seemingly peculiar feature exhibited 
by the $pp$ invariant mass distribution. Thereby we will show that the available data on 
$pp \rightarrow pp\eta$, including the invariant mass distribution, can essentially be 
understood in terms of $S$- and $P$-wave amplitudes. This result suggests that the 
properties of the $S_{11}(1535)$ nucleon resonance can be studied here in terms of only 
a few partial-wave amplitudes.
We start by considering the possible partial-wave states in $pp\rightarrow pp\eta$ near 
the threshold \cite{notation}. First, in this reaction, the $\eta$ meson is dominantly
produced through the excitation and de-excitation of the $S_{11}(1535)$ resonance.
Therefore we expect that the $\eta$ meson should be produced mainly in the $s$-wave. Of 
course, this is strictly true only in the rest frame of the resonance and not necessarily
in the overall c.m. frame of the system. However, near threshold, this should not make a 
significant difference. In fact, the observed $\eta$ angular distribution in the overall 
c.m. frame \cite{Roderburg} is practically isotropic. In addition, the very first 
analyzing power measurement in $pp\rightarrow pp\eta$ by the COSY-11 group \cite{Winter} 
yielded rather small values. Indeed this observable is basically consistent with zero, 
given the relatively large uncertainties involved, and therefore consistent with pure  
$s$-wave contributions. As long as we restrict ourselves to an $\eta$ meson in the 
$s$-wave and final $NN$ state to the $S$ and $P$ waves, we have only three partial-wave 
amplitudes that can contribute to this reaction \cite{notation}: 
$^3P_0 \rightarrow\,^1S_0 s$, $^1S_0 \rightarrow\,^3P_0 s$ and $^1D_2 \rightarrow\,^3P_2 s$.
Among these, we would naively expect that the $^3P_0 \rightarrow\,^1S_0 s$ is the only 
relevant contribution near threshold. However, as mentioned above, the contribution of 
the $S$--wave alone is unable to explain the observed bump in the $pp$ invariant mass 
distribution. 

One plausible explanation may be attributed to effects from the $\eta N$ FSI. Indeed there 
are already strong indications from the total production cross sections that the $\eta N$ 
FSI may play an important role in the reaction $NN\to NN\eta$ near threshold. For both 
$pp$ and $pn$ induced $\eta$ productions one has observed that there is an enhancement of 
the production cross sections for very small excess energies that cannot be explained by 
the $NN$ FSI effects alone \cite{Calend,Moskal0}. 
However, those effects seem to be confined to an excess energy range of up to at most 20 
MeV from the threshold. Thus, one would expect that the $\eta N$ FSI effects should 
have an influence on the invariant mass spectrum measured at the lower energy 
of $Q$ = 15 MeV. It is, however, unlikely that such effects should still be so important 
at $Q$ = 41 MeV. A proper inclusion of the $\eta N$ FSI calls for solving the Faddeev 
equation in the three-particle continuum which is technically very involved. A rough 
estimate suggests that a rather strong $\eta N$  
interaction would be needed to reproduce the data at the higher energy \cite{Baru_pc}, 
which is difficult to be reconciled with other information about the $\eta N$ 
interaction. Therefore, we seek an alternative explanation based largely on the 
observation that the shape of the bump can be reproduced by folding $p'^2$ with the 
available phase space. Here, $p'$ denotes the relative momentum in the final $pp$ system.
This suggests to us, that the bump seen in the experiment could be simply caused by the 
$pp$ $P$-wave in the final state. Admittedly, the measured final proton angular 
distribution in the overall c.m. frame is nearly isotropic \cite{Roderburg}, which could 
be seen as an evidence against large $P$-wave and higher partial wave contributions.
However, as we shall show below, there is no principal contradiction between a nearly 
isotropic proton distribution and a significant $P$-wave fraction in the invariant mass 
spectrum. 

Let us now make some general remarks about the structure of the reaction amplitude for 
$pp\rightarrow pp\eta$. In what follows, we shall assume that the $\eta$ is in an $s$ 
wave relative to the (final) $pp$ system and that the final protons are in a relative $S$
and/or $P$ state. Since the $\eta$ meson is an isoscalar pseudoscalar particle, it 
follows immediately that the orbital angular momentum of the $pp$ system has to change in
the transition from the initial to the final state and consequently, due to the Pauli 
principle, the total spin also has to change. Thus, the most general form of the reaction 
matrix (involving even angular momenta of $\eta$) can be written as
\begin{equation}
{\cal M} = \left(\vec A P_{S'=0} - \vec B P_{S'=1}\right) \cdot 
{\frac {1}{2}} \left(\vec \sigma_1 - \vec \sigma_2\right)\ ,
\label{mdef}
\end{equation}
where $P_{S'=0,1}$ stands for the total spin singlet and triplet projection operator as 
the total spin of the two protons in the final state, $S'$, takes the value $S'=0$ and 
$S'=1$, respectively. $\vec \sigma_i$ denotes the Pauli spin matrix acting on each of the 
two interacting protons, $i=1$ and $2$. In terms of the Pauli spin matrices, we have 
$P_{S'=0}=(1-\vec\sigma_1\cdot\vec\sigma_2)/4$ and 
$P_{S'=1}=(3+\vec\sigma_1\cdot\vec\sigma_2)/4$. We, then, may write, 
$P_{S'} (\vec \sigma_1 - \vec \sigma_2)/2 = \left[(\vec \sigma_1 - \vec \sigma_2) - 
(-)^{S'}i(\vec\sigma_1 \times \vec\sigma_2) \right]/4$, which, up to an irrelevant phase, 
is identical to the structure given in Ref.~\cite{Bernard}. The vectors $\vec A$ and 
$\vec B$ in Eq.~(\ref{mdef}) may be constructed from the momentum 
vectors available in the system, e.g., the relative momenta of the two protons in the 
initial state $\vec p$ and in the final state $\vec p \, '$. Since we restrict ourselves 
to $S$ and $P$ waves for the outgoing $pp$ system we may write
\begin{eqnarray}
\vec A = \alpha \hat p \ , \ \ \
\vec B = \beta \vec p \, ' + 
\gamma (\vec p \, '-3\hat p(\hat p \cdot \vec p \, ')) \ .
\label{ampsdef}
\end{eqnarray}
Here the amplitudes $\alpha$, $\beta$, and $\gamma$ correspond to the transitions 
$^3P_0 \rightarrow\,^1S_0 s$, $^1S_0 \rightarrow\,^3P_0 s$, and 
$^1D_2 \rightarrow\,^3P_2 s$, respectively. Note that we pulled out the linear momentum 
dependence, characteristic of $P$--waves, from the corresponding partial-wave amplitudes.
The amplitude $\alpha$ has a strong dependence on the relative energy of the $pp$ system 
in the final state reflecting the strong $pp$ FSI in the $^1S_0$ state. The amplitudes 
$\beta$ and $\gamma$ also depend on the relative energy of the $pp$ system in the final 
state; however, their energy dependence is much weaker than that of $\alpha$ due to the 
much weaker $pp$ FSI in the $^3P_0$ and $^3P_2$ states as compared to the $^1S_0$ state.

From Eq. (\ref{mdef}) explicit expressions for any observable follow directly. E.g., we 
find
\begin{eqnarray}
\nonumber
\frac{d\sigma}{d\Omega}\phantom{A_j} &=& |\vec A|^2+|\vec B|^2 \ , \\
\nonumber
\frac{d\sigma}{d\Omega}A_j &=& i(\vec A \, ^* \times \vec A)_j \ , \\
\frac{d\sigma}{d\Omega}C_{ij} &=& \delta_{ij}\left(|\vec A|^2-|\vec B|^2
\right)-2\mbox{Re}\left(A^*_iA_j\right) \ ,
\label{obsgen}
\end{eqnarray}
where $A_j$ denotes the analyzing power and $C_{ij}$ the spin correlation function.
Inserting the expressions of Eq.~(\ref{ampsdef}) into Eq.~(\ref{obsgen}) we get
\begin{eqnarray}
\nonumber
\frac{d\sigma}{d\Omega}\phantom{A_j} &=& |\alpha|^2+ p' \, ^2 \left[|\beta + 
\gamma|^2+3x^2(|\gamma|^2-2\mbox{Re}(\beta \gamma^*))\right] \ , \\
\nonumber
\frac{d\sigma}{d\Omega}A_j &=& 0 \ , \\
\frac{d\sigma}{d\Omega}C_{xx} &=& |\alpha|^2- p' \, ^2 \left[|\beta + 
\gamma|^2+3x^2(|\gamma|^2-2\mbox{Re}(\beta \gamma^*))\right] \ ,
\label{obs}
\end{eqnarray}
where we introduced $p'x = \vec p \, ' \cdot \hat p$. We note that partial-wave 
amplitudes with even and odd final $pp$ relative orbital angular momenta cannot interfere
with each other due to the Pauli principle.
 
Let us recall at this stage that the proton angular distribution seen in the experiment 
is isotropic \cite{Roderburg}. From the above equations we can see immediately that there
are two possible scenarios for achieving such an isotropic distribution in the presence 
of significant $pp$ $P$--wave contributions, namely 
\begin{itemize}
\item[1)] 
Dominant contributions from the transitions $^3P_0 \rightarrow\ ^1S_0 s$ and 
$^1S_0 \rightarrow\ ^3P_0 s$, but negligible contributions from 
$^1D_2 \rightarrow\ ^3P_2 s$ ($\gamma \approx 0$).
\item[2)] 
Contributions from all three transitions, $^3P_0 \rightarrow\ ^1S_0 s$, 
$^1S_0 \rightarrow\ ^3P_0 s$ and $^1D_2 \rightarrow\ ^3P_2 s$, where the latter two 
interfere destructively ($|\gamma|^2 \approx 2 {\rm Re} (\beta\gamma^*)$).
\end{itemize}
  
Obviously, the observables given in Eq.~(\ref{obs}) do not allow one to distinguish 
between the two scenarios and, consequently, there is no model independent way to extract
the two $pp$ $P$-wave amplitude contributions ($\beta$ and $\gamma$) from these 
observables. To do that, one would need observables depending also on the spin of the 
final $pp$ state, such as spin transfer coefficients. Note, however, that the combination
$d\sigma/d\Omega\,(C_{xx}+1)$ depends only on the amplitude $\alpha$. Hence, a 
measurement of this observable would determine, in a model independent way, the $pp$ 
$S$--wave contribution ($\alpha$) in the final state. Similarly, a measurement of 
$d\sigma/d\Omega\,(C_{xx}-1)$ would confirm the presence of $pp$ $P$--waves in the final 
state. It should also be stressed that $C_{xx}$ by itself is already very interesting. 
Here the two angular independent terms (first two terms in the last equation of 
(\ref{obs})) have opposite signs and thus tend to cancel each other. Consequently, this 
observable should be rather sensitive to the angular dependent term.

We now turn our attention to the results of a model for $pp\rightarrow pp\eta$. In 
Ref.~\cite{Nak1} we have presented a relativistic meson-exchange model for $\eta$ 
production in $NN$ collisions. It treats $\eta$ production in the Distorted Wave 
Born Approximation and includes both the $NN$ FSI and initial state interaction (ISI), 
the latter through the approximate procedure proposed in Ref.~\cite{ours}. While this 
model yields a satisfactory description of the near-threshold cross section data (for 
$pp\to pp\eta$ and for $pn\to pn\eta$) it fails to reproduce the recently measured 
invariant mass distributions. It should be stressed that the main objective of the 
present model calculation is not to achieve an accurate description of the existing 
data but rather to verify whether the model of Ref.~\cite{Nak1} can be modified so 
as to comply with the major features exhibited by the new data \cite{Roderburg,Moskal} 
as discussed above. 

In the development of a variant of the model \cite{Nak1} we have restricted ourselves to 
modifications of the $vNN^*$ vertex and the mixing parameter $\lambda$ in the $\pi NN^*$ 
and $\eta NN^*$ vertices. Here, $v$ stands for either the $\rho$- or $\omega$-meson and 
$N^*$ is the $S_{11}(1535)$ resonance. We also use the Paris $NN$ T-matrix \cite{Lacomb} 
as the $pp$ FSI; the Coulomb interaction is fully accounted for as described in 
Ref.~\cite{Nak2}. Everything else was kept unchanged. In contrast to Ref.~\cite{Nak1}, in 
the present work we have chosen a more general gauge invariant Lagrangian \cite{Riska} for 
the $vNN^*$ coupling
\begin{subequations}
\begin{eqnarray}
{\cal L}^{(\pm)}_{\omega NN^*}(x) & = &
\left(\frac {g_{\omega NN^*}} {m_{N^*}+m_N}\right)
\bar \psi_{N^*}(x) \gamma_5 \left\{
\left[ \gamma_\mu \frac {\partial^2}{m_{N^*}+m_N} - i\partial_\mu + 
\kappa_\omega\sigma_{\mu\nu} \partial^\nu \right]
\omega^\mu(x) \right\} \psi_N(x) \nonumber \\ 
& + & h.c.  \ ,
\label{NR12omega} \\
{\cal L}^{(\pm)}_{\rho NN^*}(x) & = & 
\left(\frac {g_{\rho NN^*}} {m_{N^*}+m_N}\right)
\bar \psi_{N^*}(x) \gamma_5 \left\{
\left[ \gamma_\mu \frac {\partial^2}{m_{N^*}+m_N} -i\partial_\mu + 
\kappa_\rho\sigma_{\mu\nu} \partial^\nu \right]
\vec \tau \cdot \vec \rho^{\,\mu }(x) \right\}\psi_N(x) 
\nonumber \\
& + & h.c.  \ ,
\label{NR12rho}
\end{eqnarray}
\label{NR12M}
\end{subequations}
where $\omega^\mu(x)$, $\vec\rho^{\,\mu}(x)$, $\psi_N(x)$ and $\psi_{N^*}(x)$ denote the 
$\omega$, $\rho$, nucleon and spin-1/2 $N^*$ (=$S_{11}(1535)$) resonance fields, 
respectively. $m_{N^*}$ denotes the mass of the nucleon resonance. The coupling constants
$g_{vNN^*}$ and $\kappa_v$ were considered to be free parameters in the calculation and 
have been adjusted to reproduce (globally) the available data, including the $pp\eta$ and
$pn\eta$ total cross sections. The $\gamma_5\gamma_\mu$ coupling at the $vNN^*$ vertex 
was necessary to achieve a reasonable fit to the data. The values obtained are:
$g_{\rho NN^*}/(m_N^*+m_N)$ = 9.5 [$fm$], $\kappa_\rho$ = -5.3 and
$g_{\omega NN^*}/(m_N^*+m_N)$ = 6.0 [$fm$], $\kappa_\omega$ = 3.8. In addition to the 
$vNN^*$ vertex given above, we have also chosen the pseudoscalar-pseudovector mixing 
parameter to be $\lambda=0.7$ at the $\pi NN^*$ and $\eta NN^*$ vertices \cite{Nak1}. All 
other parameter values are the same as given in Ref.~\cite{Nak1} corresponding to the 
case of pseudoscalar meson dominance. We refer to Ref.~\cite{Nak1} for further details of
the model.

Results for the $pp$ invariant mass distribution based on different partial wave 
contributions are shown in Fig.~\ref{fig:mpp} together with the recent data at the excess 
energies of $Q=15$ \cite{Roderburg,Moskal} and $41$ MeV \cite{Roderburg}. The full results 
are denoted by solid curves. Hereafter, these correspond to the calculations performed 
using the plane-wave basis without a partial wave decomposition and, as such, they include 
all partial waves. As is evident from the dashed curves, the observed peak in the region 
$m_{pp}^2 \sim (2m_N)^2$ is due to the strong $pp$ FSI in the $^1S_0$ state. The observed 
bump in the higher $m_{pp}$ region is largely due to the $^3P_0 s$ final state (dash-dotted 
curves). The contribution from the $^3P_2 s$ state (dotted curves) is very small. 
Contributions from other partial-wave states (mainly $^3P_2,^3F_2 \rightarrow ^1S_0d$) 
are relatively small at $Q=41$ MeV and practically negligible at $Q=15$ MeV. Thus, the 
present model is in line with the scenario (1) discussed above. Overall, the 
shape of the $pp$ invariant mass distribution exhibited by the data is nicely reproduced. 
However, the model tends to overestimate the data close to the maximum value of $m_{pp}^2$ 
at $Q=41$ MeV. In principle, this discrepancy might be due to the $p\eta$ FSI which is not 
explicitly accounted for in our model \cite{Nak1}. However, in order to reduce the predicted 
value, one needs a repulsive $p\eta$ FSI which seems to be in contradiction with all other 
evidence of $p\eta$ FSI effects in meson production \cite{Moskal0,xxx}. Moreover, no such 
discrepancy is seen at $Q=15$ MeV where the effect of the $p\eta$ FSI should be even larger. 
Further investigation is required to resolve this issue. 

It is important to note that the relative strength of the different partial-wave states 
depends crucially on the details of the model, and that means, specifically, on the 
excitation mechanism of the $S_{11}(1535)$ resonance in the present case. In fact, the 
measured $pp$ invariant mass distributions can be described with the same quality as 
shown in Fig.~\ref{fig:mpp} with the $^3P_2 s$ state contribution dominating over the 
$^3P_0 s$ state contribution. Such a scenario can easily be achieved in our model by a 
proper adjustment of the coupling constants at the $vNN^*$ vertex appearing in the 
underlying Lagrangians (Eq.~(\ref{NR12M})). However, the resulting proton angular 
distributions are then much more pronounced and, consequently, in disagreement with the 
experimental evidence \cite{Roderburg}. 

Although significant $pp$ $P$-waves in the final state seem to be necessary for 
reproducing the $pp$ invariant mass distribution, it is important to note that the energy
dependence of the total cross section near the threshold energy region is basically 
reproduced by the $pp$ FSI in the $^1S_0$ state folded with the phase space. E.g., the 
model developed by V. Baru et al. \cite{Baru} reproduces nicely the energy dependence of 
the total cross section from threshold up to $Q\sim 50$ MeV with $S$--wave contributions 
alone. Results of the present model for the total cross section are shown in 
Fig.~\ref{fig:xsc}. Comparing the curves for the $^1S_0 s$ (long-dashed) and $^1S_0 s + 
^3P_0 s$ (dash-dotted) partial waves, one realizes that the onset of the $^3P_0 s$ final 
state occurs at a fairly low excess energy and that its contribution becomes increasingly
important with increasing energy. This feature is a direct consequence of the requirement
of reproducing the $pp$ invariant mass distribution. However, as a result, the model now 
underpredicts significantly the data for energies close to threshold. The thin dashed 
curve corresponds to the $^1S_0 s$ contribution multiplied by an arbitrary factor of 3. 
This clearly shows that the total cross section data in the low energy region favor a 
larger contribution of the $^1S_0 s$ final state than is predicted by our model. Whether 
one is able to reconcile these seemingly contradictory properties within a consistent 
theoretical model remains to be seen. In any case, one should keep in mind that the 
$\eta N$ FSI, which is not included in the present model calculation, should enhance the 
$\eta$ $s$-wave contribution near threshold to some degree \cite{gar}. 

Fig.~\ref{fig:eta_ang} shows the differential cross sections as a function of the $\eta$ 
emission angle in the overall c.m. frame for two excess energies. The data from 
Ref.~\cite{Roderburg} are basically isotropic, indicating a dominant $\eta$ $s$-wave 
contribution. The theoretical results are normalized to the total cross section (obtained
by integrating the differential cross section data) in order to facilitate a proper 
comparison of their angular dependence with the experiment. At $Q=15$ MeV the 
normalization factor is about 2.7, while at $Q=41$ MeV, it is about 0.9. The dashed curves 
correspond to the $s$-wave contribution, the dash-dotted curves to the $s+p$ waves, and the
dotted curves to the $s+p+d$ waves. The last are practically indistinguishable from the
corresponding full results which are denoted by the solid lines. As expected after the 
discussion above, the angular distribution is given primarily by the $s$-wave contribution, 
with a small contribution from higher partial waves provided mainly by the $d$-wave. 

In Fig.~\ref{fig:p_ang}, the proton angular distributions in the overall c.m. frame are 
shown together with the data from Ref.~\cite{Roderburg}. Here again the model predictions
are normalized to the data using the same factor mentioned above. Evidently, at $Q=15$ 
MeV the result is isotropic; it is dominated by the $^1S_0 s$ (dashed curve) state 
followed by the $^3P_0 s$ (dash-dotted curve) state. Contributions from other partial 
waves are practically negligible. At $Q=41$ MeV there is a small contribution from
the $^3P_2 s$ (dotted curve). Obviously, the destructive interference with the $^3P_0 s$ 
state canceling the angular dependence (see Eqs.({\ref{obs})) is incomplete resulting in
a noticeable angular dependence which, however, is still compatible with the experiment. 
The difference between the dotted and solid curves is due to a small contribution from 
the $^3P_2\rightarrow ^1D_2s$ plus $^3P_2,^3F_2\rightarrow ^1S_0d$ amplitudes. 

Fig.~\ref{fig:eta_Ay} shows the prediction for the analyzing power as a function of the 
$\eta$ emission angle in the overall c.m. frame. The data are from Ref.~\cite{Winter}. 
The dashed curves correspond to the $\eta$ $s+p$ wave contributions while the dash-dotted 
curves to the $s+p+d$ waves. The solid lines are the full results. We note that the 
$s$-wave contribution alone yields $A_y \equiv 0$ (see Eqs.({\ref{obs})), so that any 
non-vanishing result must 
necessarily involve higher partial waves. Furthermore, judging from the shape exhibited by 
the analyzing power, the present model yields a vector meson dominance over the pseudoscalar 
meson in the excitation mechanism of the $S_{11}(1535)$ resonance as discussed in 
Ref.~\cite{Nak1}. Although the data indicate some contribution from partial waves higher 
than the $s$-wave, they are not sufficiently accurate to make a definitive statement as to 
the size of their contribution.

In Fig.~\ref{fig:p_Cii} we present predictions for the spin correlation function $C_{xx}$
at $Q=41$ MeV as a function of the proton angle in the overall c.m. frame. As can be seen
from Eq.~(\ref{obs}), the $^3P_0\rightarrow\,^1S_0 s$ partial wave alone leads to a 
constant value of $C_{xx}=1$. Adding the $^1S_0\rightarrow\,^3P_0 s$ contribution (dashed
curve) yields a small value of $C_{xx} \sim -0.3$. Including also the 
$^1D_2\rightarrow\,^3P_2 s$ contribution one obtains the result represented by the dotted
line. (In this context note that the $^1S_0\rightarrow\,^3P_0 s$ as well as the 
$^1D_2\rightarrow\,^3P_2 s$ contributions alone would give rise to $C_{xx}=-1$,
cf. Eq.~(\ref{obs})). Other partial-wave contributions do not change $C_{xx}$ 
qualitatively as is evident from the full result (solid curve). Thus, in our model, the 
cancellation of the $|\alpha |^2$ term and the $|\beta + \gamma|^2$ in Eqs. (\ref{obs}) 
is almost complete! This strongly enhances the relative importance of the angular 
dependent term in $C_{xx}$. Note that $C_{xx}=C_{yy}$ for all of the three partial-wave 
contributions discussed explicitly above (c.f. Eqs. (\ref{obsgen})).

In Fig.~\ref{fig:mpx} predictions for the $p\eta$ invariant mass distribution are shown 
together with the data \cite{Roderburg}. Again, the dominant contributions are from the 
$^1S_0 s$ (dashed curves) and $^3P_0 s$ (dot-dashed curves) final states; the contribution 
from the $^3P_2 s$ state (dotted curves) is negligible. At $Q=41$ MeV one sees also
some effects from other partial-wave states arising mainly from the 
$^3P_2,^3F_2 \rightarrow ^1S_0d$ amplitude. The overall shape of the measured invariant 
mass distribution is reproduced. The observed discrepancies in the details, especially at
$Q=41$ MeV, are not easy to understand in view of the nice agreement between calculated 
and measured $pp$ invariant mass distributions. 

Summarizing our results, we have shown that the currently available data on $\eta$ 
production in $pp$ collisions near the threshold energy can be largely understood in 
terms of a 
few $S$- and $P$-wave amplitudes. For a completely model-independent extraction of the 
relevant amplitudes, however, observables independent of those presently available are 
required. In this connection, the spin correlation function, either $C_{yy}$ or $C_{xx}$,
is suited to further constrain the $^3P_0 s$ and $^3P_2 s$ final state contributions. In 
any case, the final $pp$ $P$-wave contribution is crucial for explaining the measured 
invariant $pp$ mass distribution, especially, at $Q=41 MeV$. Our model calculations 
show that, the dominant amplitudes are $^3P_0\rightarrow\ ^1S_0 s$ and 
$^1S_0\rightarrow\ ^3P_0 s$. It should be stressed that in order to quantify the role 
of the $\eta N$ interaction in $pp\to pp\eta$ it is important to first understand the 
role of higher $NN$ partial waves.

Finally, we note that the present work illustrates the possibility of using meson 
production processes in $NN$ collisions to study the properties of nucleon resonances in 
terms of a few partial-wave amplitudes. In particular, the present model prediction for 
the relevant partial-wave amplitudes depends very sensitively on the details of the model
and especially to the excitation mechanism of the $S_{11}(1535)$ resonance. This offers 
an excellent opportunity to study some of the properties of the $S_{11}(1535)$ resonance 
using the $\eta$ meson production reaction in $NN$ collisions which would not be possible
to investigate in more basic reactions such as $\gamma + N \rightarrow \eta + N$ and
$M + N \rightarrow \eta + N$.

\vskip 0.5cm
{\bf Acknowledgement:}
The authors would like to acknowledge many fruitful discussions with V. Baru, and W. G. 
Love. The authors also thank W. G. Love for a careful reading of this manuscript. This 
work is supported by COSY grant No 41445282(COSY-58).

\vfill \eject

\newpage
\vglue 0.5cm
\begin{figure}
\vspace{14.5cm}
\includegraphics{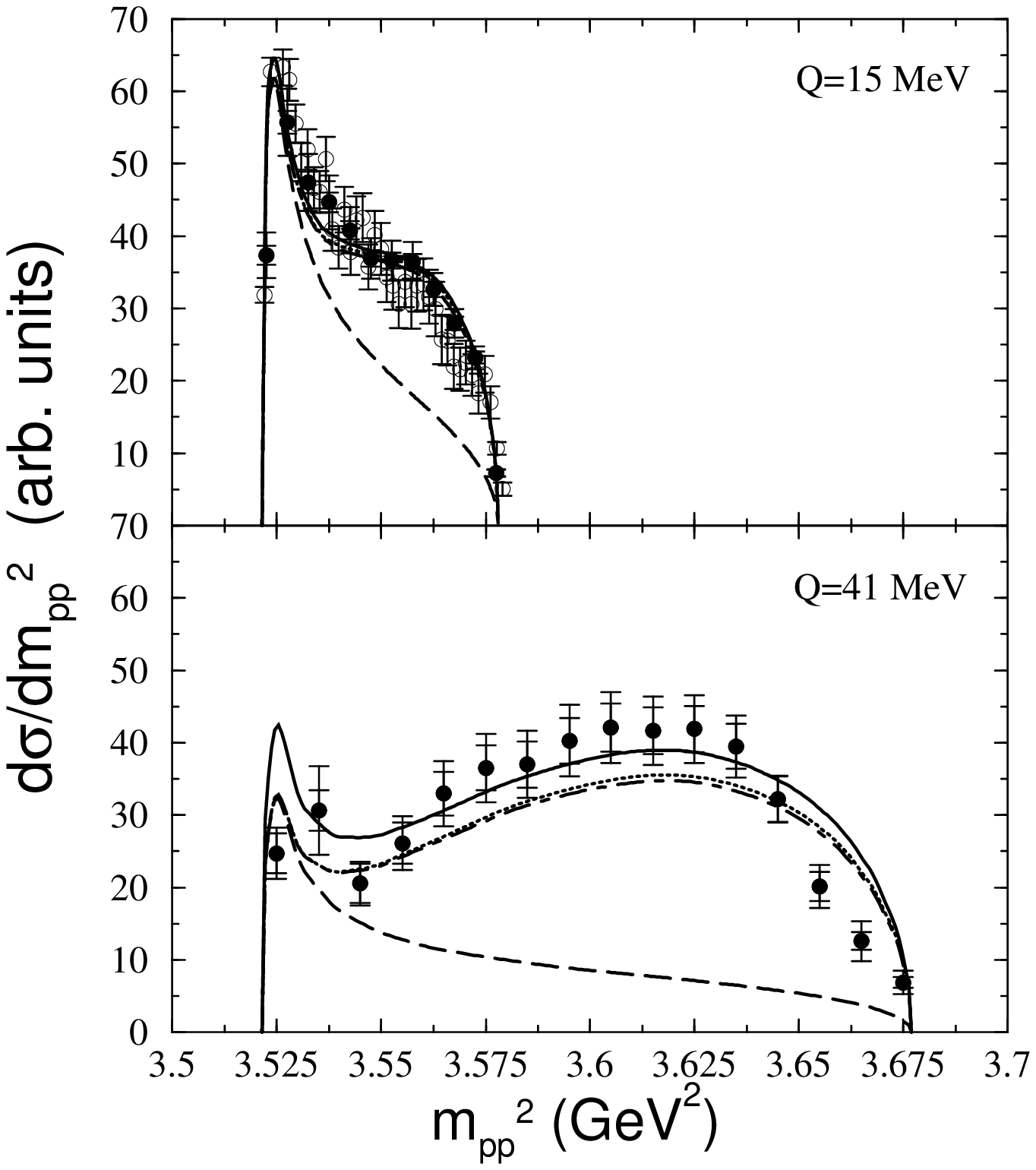}
\caption{Final $pp$ invariant mass distribution in $pp\rightarrow pp\eta$ as a function 
of invariant mass squared, $m_{pp}^2$, at an excess energy of $Q=15$ MeV (upper panel) 
and $Q=41$ MeV (lower panel). The dashed (dash-dotted) curves correspond to the $^1S_0 s$ 
($^1S_0 s + ^3P_0 s$) final state contribution. The dotted curves correspond to 
$^1S_0 s + ^3P_0 s + ^3P_2 s$; its is indistinguishable from the solid curve in the upper
panel. The solid curves are the full results. The data are from 
Ref.~\protect\cite{Roderburg} (filled circle) and Ref.~\protect\cite{Moskal} (open circle). 
The latter have been normalized by an arbitrary factor of 0.66 in order to facilitate the 
comparison of the shape with the former data and the present model prediction.}
\label{fig:mpp}
\end{figure}
\newpage
\vglue 0.5cm
\begin{figure}
\vspace{14.5cm}  
\includegraphics{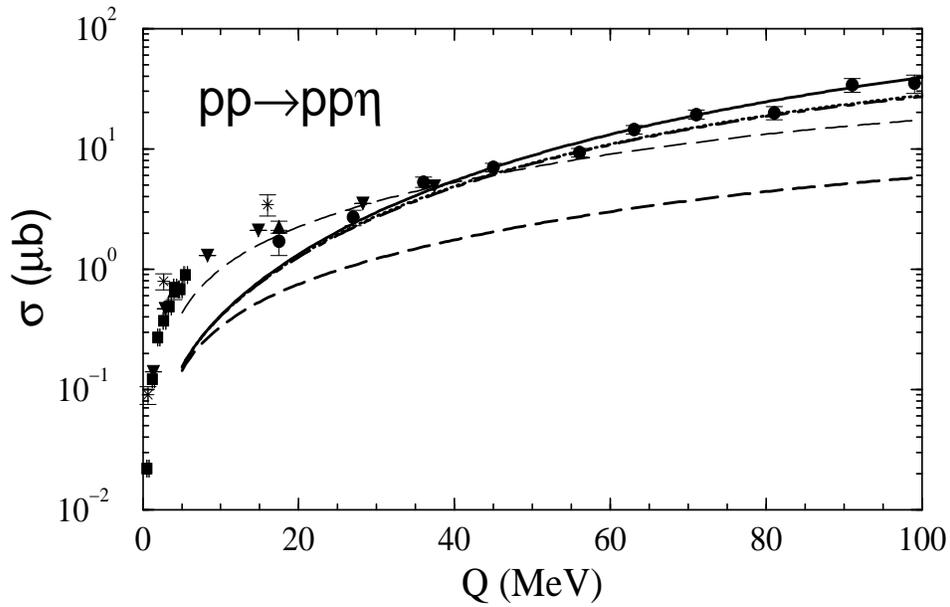}
\caption{Total cross section for the reaction $pp\rightarrow pp\eta$ as a function of the
excess energy $Q$. The solid curves represent the full results. The dashed (dash-dotted) 
curve corresponds to the $^1S_0 s$ ($^1S_0 s + ^3P_0 s$) final-state contribution and the
dotted curve to the $^1S_0 s + ^3P_0 s + ^3P_2 s$ contribution. The thin dashed curve is 
the $^1S_0 s$ contribution multiplied by an arbitrary factor of 3. The data are from 
Ref.~\protect\cite{EXP}.}
\label{fig:xsc}
\end{figure}
\newpage
\vglue 0.5cm
\begin{figure}
\vspace{15cm}
\includegraphics{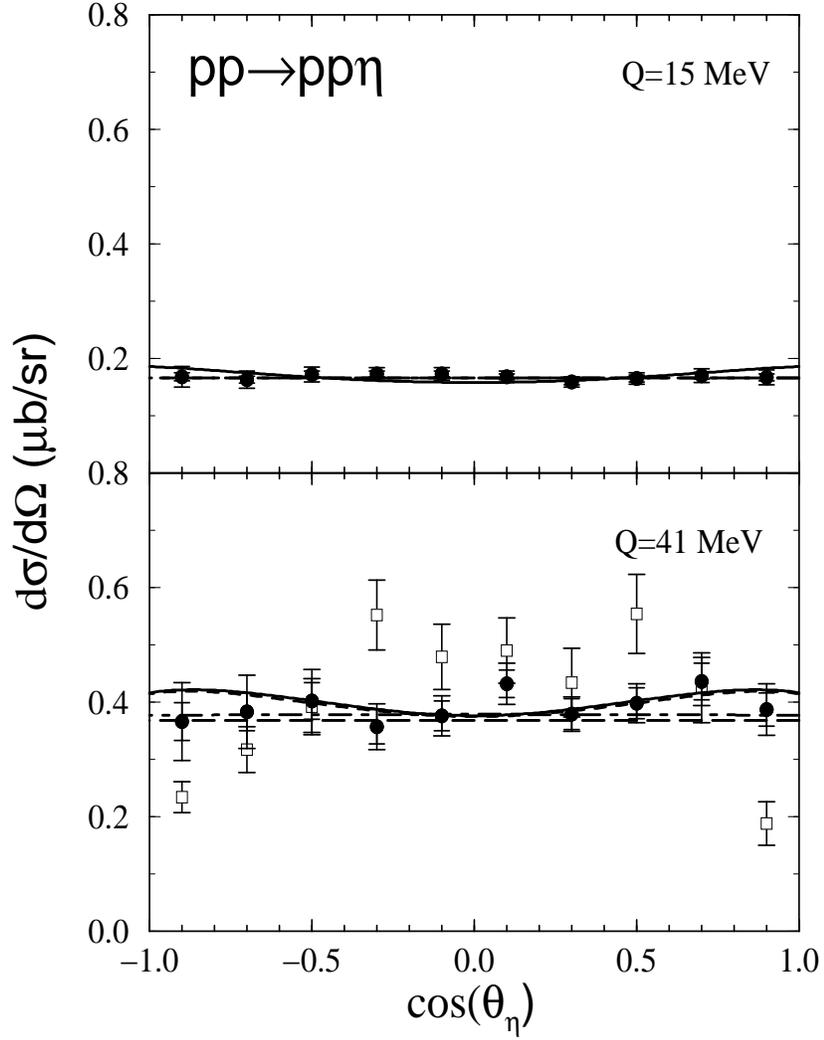}
\caption{Angular distribution of the emitted $\eta$ meson in the c.m. frame of the total 
system at an excess energy of $Q=15$ MeV (upper panel) and $Q=41$ MeV (lower panel). The 
dashed (dash-dotted) curves correspond to the $\eta$ meson $s$-wave ($s+p$-wave) 
contribution. The dotted curves correspond to the $s+p+d$-wave contributions which are 
practically indistinguishable from the corresponding full results represented by solid 
curves. The data are from Refs.~\protect\cite{Calena,Roderburg}.}
\label{fig:eta_ang}
\end{figure}
\newpage
\vglue 0.5cm
\begin{figure}[ht]
\vspace{15cm}
\includegraphics{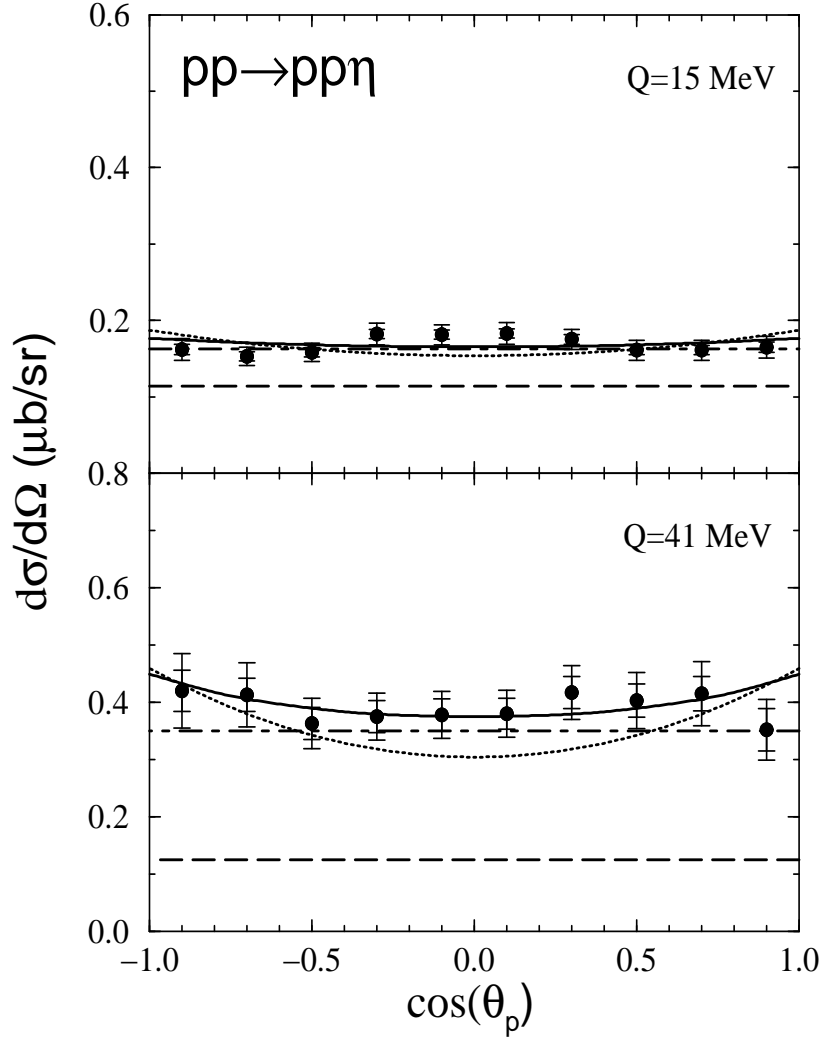}
\caption{Angular distribution of the proton in the final state in the c.m. frame of the 
total system at an excess energy of $Q=15$ MeV (upper panel) and $Q=41$ MeV (lower 
panel). The dashed (dash-dotted) curves correspond to the $^1S_0 s$ ($^1S_0 s + ^3P_0 s$)
final-state contribution. The dotted curve corresponds to the $^1S_0 s + ^3P_0 s + 
^3P_2 s$ contribution and the solid curve represents the full result. The data are from 
Ref.~\protect\cite{Roderburg}.}
\label{fig:p_ang}
\end{figure}
\newpage
\vglue 0.5cm
\begin{figure}[ht]
\vspace{15cm}
\includegraphics{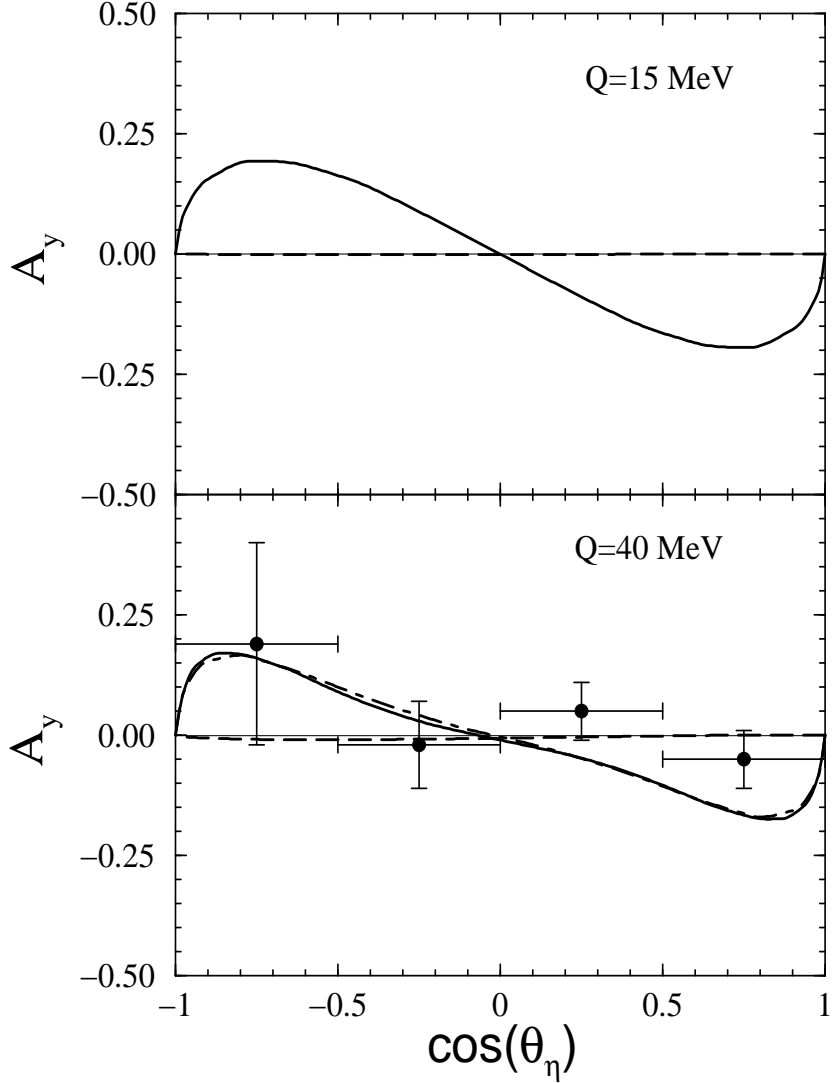}
\caption{Analyzing power for the reaction $pp\rightarrow pp\eta$ as a 
function of $\eta$ emission angle in the c.m. frame of the total 
system at an excess energy of $Q=15\ MeV$ (upper panel) and $Q=41\ MeV$ 
(lower panel). The dashed (dash-dotted) curves correspond to the partial 
waves contributions with $l'\le 1$ ($l'\le 2$). The solid curves represent the 
full results. The data are from Ref.~\protect\cite{Winter}. }
\label{fig:eta_Ay}
\end{figure}
\newpage
\vglue 0.5cm
\begin{figure}[ht]
\vspace{15cm}
\includegraphics{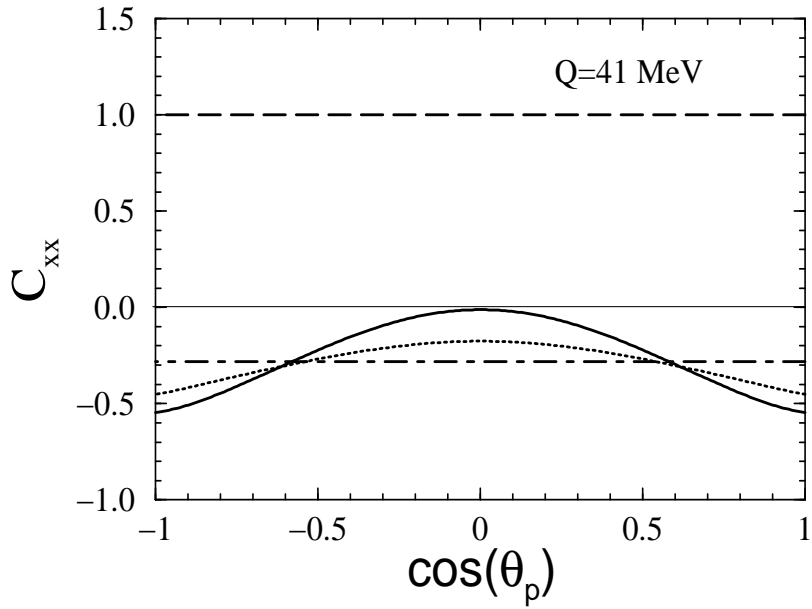}
\caption{Predicted spin correlation function $C_{xx}$ for the reaction $pp\rightarrow 
pp\eta$ as a function of final proton angle in the overall c.m. frame at an excess energy
of $Q=41$ MeV. The dash (dash-dotted) curve corresponds to the $^1S_0 s$ ($^1S_0 s 
+ ^3P_0 s$) final state contribution. The dotted curve corresponds to the $^1S_0 s + 
^3P_0 s + ^3P_2 s$ contribution and the solid curve represents the full result. }
\label{fig:p_Cii}
\end{figure}
\newpage
\vglue 0.5cm
\begin{figure}[ht]
\vspace{15cm}
\includegraphics{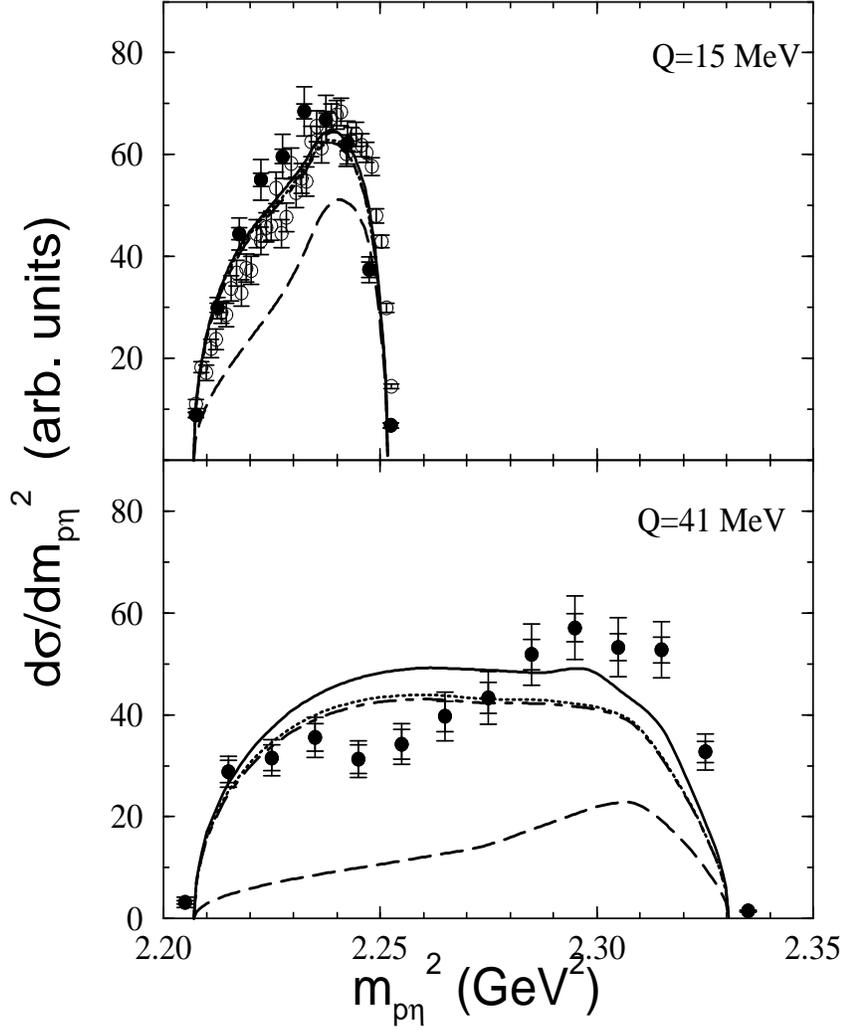}
\caption{Same as Fig.~\ref{fig:mpp} but for the $p\eta$ invariant mass distribution.
The data from Ref.~\protect\cite{Moskal} (open circle) have been normalized by an 
arbitrary factor of 0.66 in order to facilitate the comparison of the shape with 
the data from Ref.~\protect\cite{Roderburg} and the present model prediction.}
\label{fig:mpx}
\end{figure}

\end{document}